\documentclass[sigconf]{acmart}

\def\tx{\tilde{\bf X}}

\def\PI{{\bf \Pi}}
\def\N{\mathcal N}

\def\sL{\mathcal L}

\def\Thet{\boldsymbol\Theta}
\def\thet{{\boldsymbol\theta}}

\def\tx{\tilde{\bf X}}

\def\C{{\bf C}}
\def\Xz{{\bf X}_0}

\def\u0{{\bf u}_0}

\def\tx{{\tilde {\bf X}}}

\def\f{{\bf{f}}}
\def\v{{\bf{v}}}

\def\nk{{\bf n}_k}

\def\n{{\bf n}}

\def\N{{\mathcal N}}

\def\X{{\bf X}}

\def\x{{\bf x}}

\def\Xz{{\bf X}_0}

\def\b{{\bf b}}

\def\f{{\bf{f}}}
\def\v{{\bf{v}}}

\def\u{{{\bf u}}}

\def\X{{\bf X}}

\def\x{{\bf x}}

\def\Xz{{\bf X}_0}

\def\b{{\bf{b}}}

\def\C{{\bf{C}}}

\usepackage{amsthm}

\usepackage{amsmath}
\usepackage[utf8]{inputenc} 

\usepackage{multirow}
\usepackage{soul}

\usepackage{amssymb,mathrsfs}

\usepackage{epstopdf}
\usepackage{array}
\usepackage{makecell}
\usepackage{graphicx}
\usepackage{pbox}

\usepackage{commath}

\usepackage{color}
\usepackage{algorithm}
\newcommand{\svast}{\bBigg@{3}}

\usepackage{amsfonts}

\usepackage{algpseudocode}
\newcommand{\bal}{\begin{aligned}}
\newcommand{\eal}{\end{aligned}}
\newcommand{\beq}{\begin{equation}}
\newcommand{\eeq}{\end{equation}}

\newcommand{\argmax}{\operatornamewithlimits{argmax}}
\usepackage{booktabs,caption}

\usepackage{bm}

\algdef{SE}[DOWHILE]{Do}{doWhile}{\algorithmicdo}[1]{\algorithmicwhile\ #1}%

\AtBeginDocument{%
  \providecommand\BibTeX{{%
    \normalfont B\kern-0.5em{\scshape i\kern-0.25em b}\kern-0.8em\TeX}}}

\setcopyright{none}

\acmDOI{}

\acmPrice{}
\acmISBN{}

\begin{document}

\title{Inference of Regulatory Networks Through\\ Temporally Sparse Data}

\author{Mohammad Alali}
\email{alali.m@northeastern.edu}
\orcid{0000-0002-5458-5273}
\author{Mahdi Imani}
\email{m.imani@northeastern.edu}
\orcid{0000-0001-9570-9909}
\affiliation{%
  \institution{Northeastern University}
  \city{Boston}
  \state{Massachusetts}
  \country{USA}
}

\begin{abstract}
A major goal in genomics is to properly capture the complex dynamical behaviors of gene regulatory networks (GRNs). This includes inferring the complex interactions between genes, which can be used for a wide range of genomics analyses, including diagnosis or prognosis of diseases and finding effective treatments for chronic diseases such as cancer. Boolean networks have emerged as a successful class of models for capturing the behavior of GRNs. In most practical settings, inference of GRNs should be achieved through limited and temporally sparse genomics data. A large number of genes in GRNs leads to a large possible topology candidate space, which often cannot be exhaustively searched due to the limitation in computational resources. This paper develops a scalable and efficient topology inference for GRNs using Bayesian optimization and kernel-based methods. Rather than an exhaustive search over possible topologies, the proposed method constructs a Gaussian Process (GP) with a topology-inspired kernel function to account for correlation in the likelihood function. Then, using the posterior distribution of the GP model, the Bayesian optimization efficiently searches for the topology with the highest likelihood value by optimally balancing between exploration and exploitation. The performance of the proposed method is demonstrated through comprehensive numerical experiments using a well-known mammalian cell-cycle network.
\end{abstract}

\ccsdesc[500]{Mathematics of computing~Markov processes}
\ccsdesc[500]{Mathematics of computing~Maximum likelihood estimation}
\ccsdesc[300]{Mathematics of computing~Bayesian computation}
\ccsdesc[300]{Mathematics of computing~Nonparametric statistics}

\keywords{Topology Inference, Gene Regulatory Networks, Boolean Dynamical Systems, Bayesian Optimization}


\maketitle

\section{Introduction}
Gene regulatory networks (GRNs) play an important role in the molecular mechanism of underlying biological processes, such as stress response, DNA repair, and other mechanisms involved in complex diseases such as cancer. A deep understanding of these biological processes is key in diagnosing and treating these chronic diseases. Advances in high-throughput genomic and proteomic profiling technologies have provided novel platforms for studying genomics. Meanwhile, single-cell gene-expression measurements allow capturing multiple snapshots of these complex biological processes. These advances offer an opportunity for seeking systematic approaches to understand the structure of GRNs. The topology inference of GRNs is critical in systems biology since it can generate valuable hypotheses to promote further biological research.

Many techniques have been developed to identify and characterize the behavior of GRNs. The correlation-based networks have been developed to better understand the relationships among different genes by analyzing multi-modal data sets~\cite{langfelder2008wgcna}. Unfortunately, these methods cannot take advantage of the information inherent in time-series genomic data. More relevant approaches considering time components are Gaussian graphical models~\cite{vinciotti2016model,chiquet2019multiattribute}, regression analyses~\cite{dong2013inference,salleh2017multiple}, heuristic methods such as genetic algorithm and random forest~\cite{BMC1,BMC2}, information theoretical approaches~\cite{villaverde2013reverse}, and non-parametric Bayesian models~\cite{chan2017gene, BMC3}. However, these models suffer from a lack of interpretability, as they cannot directly predict the transcriptional dynamics and resolve the directionality of regulatory interactions. The ordinary differential equation (ODE) models~\cite{henriques2015reverse,yang2020overview}, non-linear differential equation models such as an S-system~\cite{ma2020inference,mandal2016reverse}, probabilistic graphical models~\cite{kotiang2020probabilistic}, and dynamic Bayesian networks~\cite{de2017inferring} are well-studied dynamical models for GRNs. The ODE and nonlinear differential equation models suffer from overfitting and lack of interpretability; the probabilistic graphical models and dynamic Bayesian networks can effectively model dynamic processes, as well as discover causal interactions, but suffer from a lack of scalability and the existence of a large number of parameters, which often cannot be properly inferred through a relatively small number of samples. The Boolean networks~\cite{shmulevich2002probabilistic,shmulevich2002boolean,ostrowski2016boolean} (e.g., probabilistic Boolean network, Boolean network with perturbation, Boolean control networks) are highly interpretable mathematical models capable of modeling the bidirectional interactions between genes that are aligned with how biologists interpret regulatory networks. However, the lack of scalability has limited their applications to small GRNs or GRNs with rich available prior knowledge.

This paper is focused on the inference of GRNs modeled with Boolean networks and addresses the following two challenges in the GRNs' inference:
\begin{itemize}
    \item {\em Large Topology Candidate Space:} The modeling consists of estimating a large number of interacting parameters, which represent the connections between genes that govern their dynamics. This requires searching over a large number of topology candidates and picking the one with the highest likelihood value given the available data. Most existing inference methods for general nonlinear models are developed to deal with continuous parameter spaces, such as maximum likelihood~\cite{ImanBrag:TSPJ2,ImanBrag:BBKFJ12,kantas2015particle,johansen2008particle}, expectation maximization~\cite{Godsill2004,Hurzeler1998,Schon2011,will2013}, and multi-fidelity~\cite{Iman:AAAIC15,ImanGh:TNNLSJ17} methods. However, these methods cannot be applied for inference over large discrete parameter spaces, such as the large topology candidate space of GRNs. In this paper, we develop a method that is scalable with respect to the number of unknown interactions, and efficiently searches over the large topology candidate space. More specifically, our proposed method enables optimal inference in the presence of a large number of unknown regulations for GRNs with a relatively small number of genes.
    \item {\em Expensive Likelihood Evaluation:} The likelihood function, which measures the probability that the available data come from each topology candidate, is often expensive to evaluate. The reasons for that are a large number of genes in GRNs, and the sparsity in the data, which require propagation of the system stochasticity across time and gene states. Given the limitation in the computational resources, evaluation of the likelihood functions for all of the topology candidates is impossible, and one needs to find the topology with the highest likelihood value with a few expensive likelihood evaluations. 
\end{itemize}


 This paper derives a scalable topology inference for GRNs observed through temporally sparse data. The proposed framework models the expensive-to-evaluate (log-)likelihood function using a Gaussian Process (GP) regression with a structurally-inspired kernel function. The proposed kernel function exploits the structure of GRNs to efficiently learn the correlation over the topologies, and enables Bayesian prediction of the log-likelihood function for all the topology candidates. Then, a sample-efficient search over topology space is achieved through a Bayesian optimization policy, which sequentially selects topologies for likelihood evaluation according to the posterior distribution of the GP model. The proposed method optimally balances exploration and exploitation, and searches for the global solution without getting trapped in the local solutions. The accuracy and robustness of the proposed framework are demonstrated through comprehensive numerical experiments using a well-known mammalian cell-cycle network.

 The remainder of this paper is organized as follows. Section 2 provides a detailed description of the GRN model and the topology inference of GRNs. Further, the proposed topology optimization framework is introduced in section 3. Section 4 presents various numerical results, and the main conclusions are discussed in section 5.

\label{sec:Intro}

\section{Preliminaries}
\label{sec:pre}
\subsection{GRN Model}
This paper employs a Boolean network with perturbation (BNp) model for capturing the dynamics of GRNs~\cite{shmulevich2010probabilistic,Iman:CIJ5,EhsanBMCGenomics}. This model properly captures the stochasticity in GRNs, coming from intrinsic uncertainty or unmodeled parts of the systems. Consider a GRN consisting of $d$ genes. The {\em state process} can be expressed as $\{\X_k; k=0,1,\ldots\}$, where $\X_k \in \{0,1\}^d$ represents the activation/inactivation state of the genes at time $k$. The gene state is updated at each discrete time through the following Boolean signal model:
\beq\label{eq-X}
    \X_k \,=\, \f(\X_{k-1})\,\oplus\,\n_k\,, 
\eeq
for $k=1,2,\ldots$, where $\n_k\in\{0,1\}^d$ is Boolean transition noise at time $k$, ``$\oplus$" indicates component-wise modulo-2 addition, and $\f$ represents the {\em network function}. The network function model for GRNs can be expressed according to the pathway diagram as:
\beq\label{eq:BN3}
    \f(\X_{k-1}) \,=\, \overline{\C\, \X_{k-1}\,+\,\b},
\eeq
where $\C$ is the connectivity matrix, $\b$ is the bias vector, and the threshold operator $\overline{\v}$ maps the positive elements of vector $\v$ to $1$ and negative elements to $0$. $\C$ is a $d\times d$ matrix, where the element in the $i$th row and $j$th column $c_{ij}$ takes one of the following three values: $c_{ij}=+1$ if there is a positive regulation (activation) from gene $j$ to gene $i$; $c_{ij}=-1$ if there is a negative regulation (inhibition) from gene $j$ to gene~$i$; and $c_{ij}=0$ if gene $j$ is not an input to gene $i$. $\b$ is a vector of size $d$ where each element of this vector takes $-1/2$ or $+1/2$. The elements of this vector are tiebreakers for zero-elements of vector $\C\X_{k-1}$;  $-1/2$ and $+1/2$ map the corresponding state value of genes to $0$ or $1$ respectively.

In (\ref{eq-X}), the noise process $\nk$ indicates the amount of stochasticity in a Boolean state process. For example, $\n_{k}(j)=1$, means that the $j$th gene's state at time step $k$ is flipped and does not follow the Boolean function. Whereas, $\n_{k}(j)=0$ indicates that this state is governed by the network function. We assume that all the $\nk$ components are independent and have a Bernoulli distribution with parameter~$(p)$, which $0\leq p < 0.5$ refers to the amount of stochasticity in each state variable (i.e., gene).  
 
\subsection{Topology Inference of Regulatory Networks}
In practice, the network function is unknown or partially known, and the unknown parameters need to be inferred through available data. The unknown information is often the elements of the connectivity matrix or bias units. Without loss of generality, we assume that $L$ elements of the connectivity matrix are unknown. Given that each element takes in values in space $\{+1,0,-1\}$, there will be $3^L$ different possible combinations for these missing interactions, denoted by $\Theta=\{\thet^1,...,\thet^{3^L}\}$. Each set of these possible parameters refers to a possible network function for the GRN (e.g., $\C^\thet$ represents the connectivity matrix associated with parameter vector $\thet$),  while only one parameter represents the true unknown system topology. Assuming that $D_{1:T}$ represents the available data, the inference process can be formulated as:
\beq\label{eq:OPT}
\thet^*=\argmax_{\thet\in\Theta} P(D_{1:T}\mid\thet),  
\eeq
where $P(D_{1:T}\mid\thet)$ is the likelihood function for the topology parameterized by $\thet$. The solution to the optimization problem, $\thet^*$ in (\ref{eq:OPT}), is known as the maximum likelihood solution. Note that without loss of generality, the proposed method, which will be described in the next section, can be applied to any arbitrary point-based estimator such as maximum aposteriori.

\section{The proposed Framework}
\subsection{Likelihood Evaluation}
\label{sec:PF}
Let $\{\x^1,\ldots,\x^{2^{d}}\}$ be an arbitrary enumeration of the possible Boolean state vectors (i.e., a GRN with $d$ components). The available data in $D_{1:T}$ can be represented using the vector $I_{1:T}=\{I_1,....,I_T\}$, where $I_k$ specifies the index associated with the $k$th state ($0$ if the the state at time step $k$ is missing). For instance, $D_{1:6}=\{\tx_2=\x^9,\tx_3=\x^3,\tx_6=\x^{11}\}$ contains the information of time steps 2, 3 and 6, and denotes that data at time steps 1, 4 and 5 are missing. In this case, the indicator vector is defined as $I_{1:6}=\{0,9,3,0,0,11\}$.

For any given model $\thet\in\Theta$, we define the predictive posterior distribution $(\PI^{\thet}_{k|k-1})$ and posterior distribution $(\PI^{\thet}_{k|k})$ of the states as:
\beq\label{eq:Pi}
 \bal
 \PI^{\thet}_{k|k-1}(i) \,&=\, P\left(\X_{k} =\x^i\, \mid \, I_{1:k-1},\thet\right)\,,\\
 \PI^{\thet}_{k|k}(i) \,&=\, P\left(\X_{k} =\x^i\, \mid \, I_{1:k},\thet\right)\,,
\eal
\eeq 
for $i=1,\ldots,2^{d}$ and $k=1,2,...$ . 

We define the {\em transition matrix} $M^\thet$ of size $2^{d} \times 2^{d}$ associated with a GRN model parameterized by $\thet$, through the following notation:
\beq\label{eq:M}
\bal
  (M^\thet)_{ij} \,&=\, P\left(\X_r= \x^i \,\,\mid\,\, \X_{r-1} = \x^j,\thet\right)\,, \\
  &=\, P\left(\n_r \,=\,\f^\thet(\x^j) \oplus \x^i \right)\,\\[5pt]
        & =\, p^{||  \overline{\C^\thet\, \x^j\,+\,\b}\, \oplus \,\x^i||_1}  (1\!-\!p)^{d-||  \overline{\C^\thet\, \x^j\,+\,\b} \,\oplus \,\x^i||_1}\,, 
\eal
\eeq
for $i,j = 1,\ldots,2^{d}$, where the second and third lines in (\ref{eq:M}) are obtained based on the GRN model in~(\ref{eq-X}).

Let $\PI^\thet_{0|0}(i)=P(\X_0=\x^i\mid\thet)$, for $i=1,...,2^d$, be the initial state distribution. If no knowledge about this distribution is available, this can be represented by $\PI^\thet_{0|0}(i)=1/2^d$, for $i=1,...,2^d$, and $\thet\in\Theta$. The predictive posterior distribution can be computed recursively as:
\beq\label{eq:Trans}
\PI^\thet_{k|k-1}=M^\thet\PI^\thet_{k-1|k-1}.
\eeq
If the data at time step $k$ is missing, the posterior distribution is equal to predictive posterior distribution, otherwise it becomes 1 for the observed state and 0 for other elements. This can be formulated as: 
\beq
\PI^\thet_{k|k}(i)=\begin{cases}
\PI^\thet_{k|k-1}(i)  \hspace{1.3cm} \text{if} \quad I_k=0,\\
\PI^\thet_{k|k-1}(I_k)=1 \hspace{0.66cm} \text{if} \quad i=I_k,\\
0 \hspace{2.5cm} \text{otherwise}
\end{cases}
\eeq
for $i=1,\ldots,2^{d}$ and $k=1,2,...$ . 

The likelihood value in optimization problem in (\ref{eq:OPT}) can be written in logarithmic format as: 
\beq\label{eq--Li}
\bal
L(\thet)&:=\log P(D_{1:T}\mid\thet)=\log P(I_{1:T}\mid\thet)\\ 
&=\sum_{k=1}^T \log P(I_k\mid I_{1:k-1},\thet),
\eal
\eeq
where
\beq
 P(I_k\mid I_{1:k-1},\thet)=\begin{cases}
 \PI_{k|k-1}^\thet(I_k)  &\text{if }I_k\neq 0\\
1 & \text{otherwise}
\end{cases}
\eeq

The computation of the log-likelihood value for any given topology can be huge due to the large size of the transition matrices with $2^{2d}$ elements. The computational complexity of log-likelihood evaluation is of order $O(2^{2d} T)$, where $T$ is the time horizon. This substantial computational burden (especially in systems with a large number of components) is the motivation to come up with more efficient ways to solve the problem presented in (\ref{eq:OPT}).

\subsection{Bayesian Optimization for Topology Optimization}
This article proposes a Bayesian optimization approach for scalable topology inference of regulatory networks observed through temporally sparse data. The main concepts of this approach are explained in detail in the following paragraphs. \vspace{.6ex}

\noindent
\underline{\bf GP Model over the Log-Likelihood Function:}
The transition matrix ($M^{\thet}$) in (\ref{eq:Trans}) makes the log-likelihood function evaluation in (\ref{eq:OPT}) and (\ref{eq--Li}) computationally expensive, especially when dealing with large scale regulatory networks. Therefore, it is vital to come up with an efficient way of searching over the topology space. In this article, the log-likelihood function $L(.)$ is modeled using the Gaussian Process (GP) regression. The GP~\cite{rasmussen2006gaussian} is mostly defined over continuous spaces, primarily due to the possibility of defining kernel functions that model the correlation over continuous spaces. In our case, the parameters are discrete interactions (i.e., parameters of the connectivity matrix that take $+1$, $0$, or  $-1$), which prevent constructing the GP model for representing the log-likelihood function over topology space. 

This paper takes advantage of the topology structure of GRNs, encoded in connectivity matrix in (\ref{eq:BN3}), and defines the following GP model: 
\beq\label{eq:gp}
\sL(\thet)\, = {\mathcal GP}\left(\mu(\thet), k (\thet,\thet)\right),
\eeq
where $\mu(.)$ shows the mean function, and $k(., .)$ indicates the topology-inspired kernel function. Using the fact that each interacting parameter $\thet$ corresponds to a connectivity matrix $\C^\thet$, the structurally-inspired kernel function is defined as:
\beq
k(\thet,\thet')=\sigma_f^2\exp\left(-\frac{|| C^{\thet}-C^{\thet'}||^2}{l}\right),
\eeq
where $||{\bf V}||^2$ is the sum of squares of elements of ${\bf V}$, $C^\thet$ and $C^{\thet'}$ represent the connectivity matrices related to topologies $\thet$ and $\thet'$ respectively, $l$ is the length-scale, and $\sigma_f^2$ is the scale factor hyperparameters. These hyperparameters quantify how close the topologies are to each other. The more similar two topologies are (i.e., less difference in the connectivity matrices), the more they are correlated, and the kernel function value will be higher for them. While, for more distinct topologies, the kernel will have smaller values.

\begin{figure}[ht!]
\centering
\includegraphics[width=70mm]{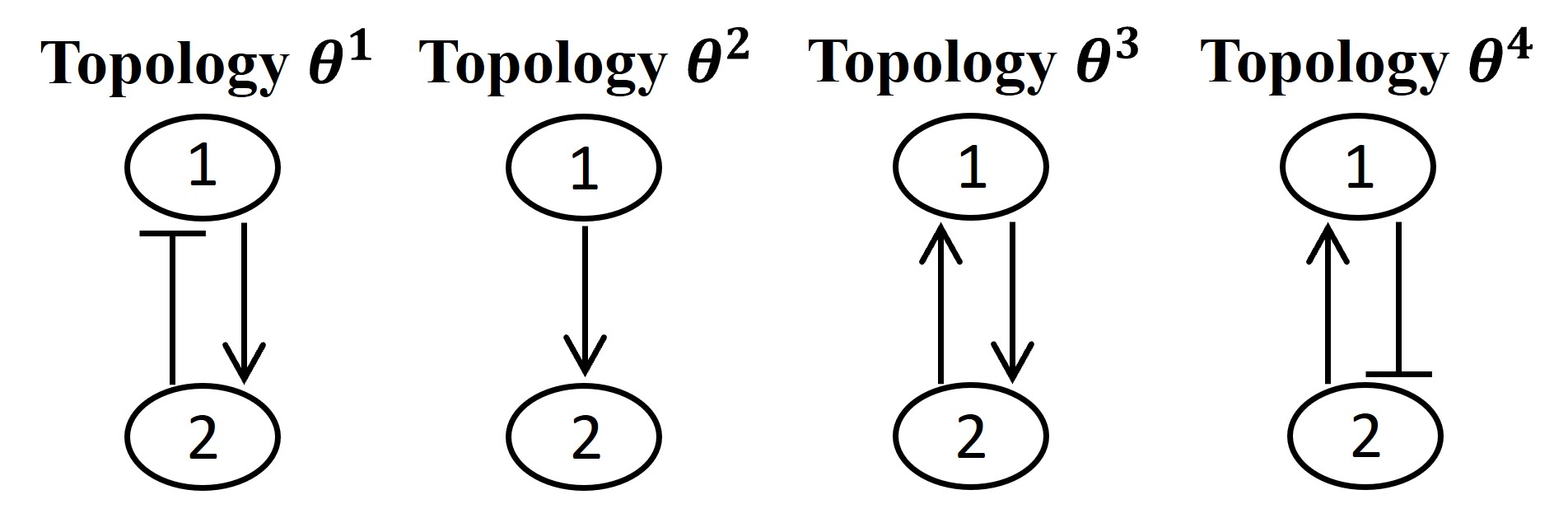}\vspace{0ex}
 \caption{An example of possible models (i.e, topologies) for a GRN with two genes.}
 \label{fig:Ex}
\end{figure} 

Figure~\ref{fig:Ex} represents an example of a few possible topologies for a GRN with two genes. These four possible topologies differ in one or two interactions. If the log-likelihood value for topology $\thet^1$ is calculated, this information can be used for predicting log-likelihood values for other topologies. The connectivity matrices for these topologies can be expressed as:
\beq
\bal
\C^{\thet^1}\!=\!\begin{bmatrix}
0 & -1\\
1 & 0\\
\end{bmatrix}, 
\C^{\thet^2}\!=\!\begin{bmatrix}
0 & 0\\
1 & 0\\
\end{bmatrix},
\C^{\thet^3}\!=\!\begin{bmatrix}
0 & 1\\
1 & 0\\
\end{bmatrix},
\C^{\thet^4}\!=\!\begin{bmatrix}
0 & 1\\
-1 & 0\\
\end{bmatrix}.
\eal\label{eq:ex}
\eeq
{The correlation between topology $\thet^1$ and all the aforementioned topologies, $\Theta^{\rm sub}=\{\thet^1,\thet^2,\thet^3,\thet^4\}$ can be expressed through the following kernel vector: 
\beq
\bal
{\bf K}_{(\thet^1,\Theta^{\rm sub})}&\!=\!\begin{bmatrix}
k(\thet^1,\thet^1)\!\! & k(\thet^1,\thet^2)\!\! &k(\thet^1,\thet^3) &\!\! k(\thet^1,\thet^4)\\ 
\end{bmatrix} \\
&=\sigma_f^2\begin{bmatrix}
1 & 1/\exp(1) &1/\exp(4) &1/\exp(16)\\ 
\end{bmatrix},
\eal
\eeq}
where the length-scale hyperparameter is assumed to be $1$. It can be seen that topology $\thet^1$ has the maximum correlation with itself, and the correlation rate decreases when we move from topology $\thet^1$ to $\thet^4$. This can also be understood in terms of the differences between the interacting parameters, expressed in the connectivity matrices in (\ref{eq:ex}). The difference between topology $\thet^1$ and topology $\thet^2$ is a missing interaction from gene 2 to gene 1, which has led to a relatively large correlation between these two topologies. The interaction from gene 2 to gene 1 is suppressive and activating in $\thet^1$ and $\thet^3$, respectively, which is expressed in smaller correlation values between topologies $\thet^1$ and $\thet^3$. Finally, $\thet^1$ and $\thet^4$ have two opposite types of interactions, leading to the smallest correlation between these two topologies. 

One possible choice for the mean functions $\mu(.)$ in (\ref{eq:gp}), which represents the prior shape of the log-likelihood function over all topologies, is the constant mean function. This mean function carries a single hyperparameter, which can be learned along with the kernel hyperparameters.  

The GP model has the capability of providing the Bayesian representation of the likelihood function across the topology space. Let ${\thet}_{1:t}= (\thet_1,\ldots,\thet_t)$  be the first $t$ samples from the parameter space (i.e., samples from the topology candidates) with the associated log-likelihood values $L_{1:t} = [L_1,..., L_t]^T$ (i.e., $L_1=L(\thet_1)$ in (\ref{eq--Li}). The posterior distribution of $\sL(\thet)$ in equation (\ref{eq:gp}) is derived as:
\beq
\sL(\thet)\mid\thet_{1:t},L_{1:t}\sim \N(\mu^t_\thet,\Sigma^t_\thet),
\eeq
where ${\mu}^t_\thet$ and ${\Sigma}^t_\thet$ are the mean and variance for a specific model $\thet\in \Theta$ respectively. These values can be obtained as: 
\beq\label{eq:fb}
\bal
{\mu}^t_\thet&=\mu({\thet})+{\bf K}_{({\thet},\thet_{1:t})}\, {\bf K}_{(\thet_{1:t},\thet_{1:t})}^{-1}\,(L_{1:t}- \boldsymbol{\mu}(\thet_{1:t})),\quad\\[1ex]
{\Sigma}^t_\thet&=\,{k}_{({\thet},{\thet})}-{\bf K}_{({\thet},\thet_{1:t})} \,{\bf K}_{(\thet_{1:t},\thet_{1:t})}^{-1}\, {\bf K}_{({\thet},\thet_{1:t})}^T\,,
\eal\vspace{0ex}
\eeq
where $\boldsymbol{\mu}(\thet_{1:t})=[\mu(\thet_1),...,\mu(\thet_t)]^T$, and
\beq
{\bf K}_{(\Thet,\Thet')}=\begin{bmatrix}
k(\thet_1,\thet'_1) & \ldots & k(\thet_1,\thet'_r)\\
\vdots & \ddots & \vdots\\
k(\thet_l,\thet'_1) & \ldots & k(\thet_l,\thet'_r)\\ 
\end{bmatrix}, 
\eeq
for $\Thet=\{\thet_1,...,\thet_l\}$, $\Thet'=\{\thet'_1,...,\thet'_r\}$.
Using the aforementioned formulation, the GP constructs the log-likelihood function as a zero-mean Bayesian surrogate model with covariance $k(.,.)$. Further, at iteration $t$, the log-likelihood function can be computed by employing the already chosen and evaluated log-likelihood values for topologies $\thet_{1:t}$, i.e., $L_{1:t}$. The uncertainty of the surrogate model will be reduced as we evaluate the likelihood function for more topologies.  

The GP hyperparameters, which consist of the hyperparameters of the topology-inspired kernel function and the mean function, can
be learned by optimizing the marginal likelihood function of the GP model at each iteration through: 
\beq
L_{1:t} \mid \thet_{1:t} \sim \N \left(\boldsymbol{\mu}(\thet_{1:t}), {\bf K}_{(\thet_{1:t},\thet_{1:t})}\right).
\eeq

\begin{algorithm*} 
\caption{The Proposed method for inference of regulatory networks through temporally-sparse data.}
\begin{algorithmic}[1]  
\State Topology space $\Theta$; data $D_{1:T}$; initialize the hyperparameters of the Gaussian process ${\mathcal GP}(\mu(.),k(.,.))$, $t=0$. \vspace{1ex}
\State Arbitrary enumeration of the possible  Boolean state vectors: $\X=[\x^1,\ldots,\x^{2^{d}}]$ \vspace{1ex}
\State Initialization: $I=\vec{0}_{1 \times T}$
\vspace{1ex}
\noindent
\For {$k = 1,2,\ldots,T$}\vspace{1.ex}
\For {$i=1,\ldots,2^d$}\vspace{1.ex}
\State $I_k =\begin{cases}
 i  &\text{if } D_k \text{ is not missing and } D_k = \X_i\\
0 &\text{if } D_k \text{ is missing}
\end{cases}$ \vspace{1ex}
\EndFor\vspace{1ex}
\EndFor\vspace{1ex}

\Repeat{}\vspace{1ex}
\Statex \hspace{3ex}\underline{\bf Sequential Topology Selection}\vspace{1ex}

      \State Pick the topology with maximum acquisition value:
$\thet_{t+1}=\argmax_{\thet\in\Theta}\alpha_t(\thet)$ --- Eq.~\ref{eq:EI} \vspace{1ex}
\Statex \hspace{3ex}\underline{\bf Log-Likelihood Computation}\vspace{1ex}
\State Initialization: $\PI^{\thet_{t+1}}_{0|0}(i) \,=\, P\left(\Xz = \x^i\mid\thet_{t+1}\right)$, for $i=1,\ldots,2^d$, $L_{t+1}=0$.  
\vspace{1ex}
\noindent
\For {$k = 1,2,\ldots,T$}\vspace{1.ex}
 \State Predictive Posterior Distribution: $\PI^{\thet_{t+1}}_{k|k-1}=M^{\thet_{t+1}}\PI^{\thet_{t+1}}_{k-1|k-1}$. 
 \vspace{1ex}
\State Posterior Distribution: $
\PI^{\thet_{t+1}}_{k|k}(i)=\begin{cases}
\PI^{\thet_{t+1}}_{k|k-1}(i)  \hspace{1.3cm} \text{if} \quad I_k=0,\\
\PI^{\thet_{t+1}}_{k|k-1}(I_k)=1 \hspace{0.66cm} \text{if} \quad i=I_k,\\
0 \hspace{2.5cm} \text{otherwise}
\end{cases}$ \vspace{1.ex}
 \State Log-Likelihood Update: $L_{t+1}=L_{t+1}\!+\!\log P(I_k\mid I_{1:k-1},{\thet_{t+1}}), \text{where } P(I_k\mid I_{1:k-1},{\thet_{t+1}})=\begin{cases}
 \PI_{k|k-1}^{\thet_{t+1}}(I_k)  &\text{if }I_k\neq 0\\
1 & \text{otherwise}
\end{cases}$ \vspace{1ex}
\EndFor\vspace{1ex}

\Statex \hspace{3ex}\underline{\bf Update the GP Model}\vspace{1ex}
      \State Update the hyperparameters of the GP according to $(\thet_{1:t+1},L_{1:t+1})$.\vspace{1.2ex}
\State $t=t+1.$\vspace{1ex}
\Until{the stopping criterion is met } \vspace{1ex} 
 \State The Inferred Topology: $\thet^*:=\thet_{i^*}, \text{ where }i^*=\argmax_{i=1,...,t} L_i\,$. \vspace{1ex}
   \end{algorithmic}\label{alg:1}
\end{algorithm*}

\noindent
\underline{\bf Sequential Topology Optimization:} The notion of efficient topology optimization is to come up with an efficient way of searching over all the topology space so that we utilize a minimum number of computationally expensive likelihood evaluations and eventually find the optimal topology, which yields the largest likelihood value.

The log-likelihood function is computationally expensive to evaluate, therefore, the sample-efficient and sequential topology selection is achieved as:
\beq\label{eq:EI}
\thet_{t+1}=\argmax_{\thet\in\Theta} \alpha_{t}(\thet).
\eeq
where $\alpha_{t}(\thet)$ represents the acquisition function in the Bayesian optimization context, which is determined over the GP model posterior at iteration $t$. Multiple acquisition functions exist in the context of Bayesian optimization (BO), and all of them try to balance the exploration and exploitation during the optimization process. Without loss of generality, in this work, we use the expected improvement acquisition function, which is defined as~\cite{mockus1978application,jones1998efficient}:
\beq\label{eq-EI}
\bal
\alpha_{t}(\thet)  &=\left(\mu_\thet^t-L_{\rm max}^t\right)\,\Phi\bigg(\left(\mu_\thet^t-L_{\rm max}^t\right)/\sqrt{\Sigma_\thet^t}\bigg)\\[-.3ex]
&\qquad\qquad\qquad+\sqrt{\Sigma_\thet^t}\,\phi\bigg(\left(\mu_\thet^t-L_{\rm max}^t\right)/\sqrt{\Sigma_\thet^t}\bigg),
\eal
\eeq
where $\phi(.)$ and $\Phi(.)$ refer to the probability density function and cumulative density function of standard normal distribution, $L_{\rm max}^t=\max\{L_1,...L_t\}$ is the maximum log-likelihood value until the latest turn, and $\mu_\thet^t$ and $\Sigma_\thet^t$ are the mean and variance of the GP model at iteration $t$ as defined in (\ref{eq:fb}).

The acquisition function in~(\ref{eq-EI}) holds a closed-form solution and requires the mean and variance of the GP model for any given topology. The next log-likelihood evaluation is carried out for topology $\thet_{t+1}$ to derive the log-likelihood value $L_{t+1}$. 
The GP model is then updated based on all the new information, defined as $\thet_{1:t+1}=(\thet_{1:t},\thet_{t+1})$ and
${L}_{1:t+1}=[L_{1:t}, L_{t+1}]^T$. In large regulatory networks where we might have a large number of unknown interactions, we can implement some heuristic optimization methods including particle swarm optimization technique~\cite{kennedy1995particle}, genetic algorithm~\cite{anderson1994genetic,whitley1994genetic}, or the breadth-first local search (BFLS)~\cite{atabakhsh1991survey} to obtain the model with the largest acquisition value. 

The proposed Bayesian topology optimization continues its sequential process over all the topology space of the regulatory networks for a fixed number of turns or until no significant change in the maximum log-likelihood value in consecutive iterations is spotted. When the optimization ends, the topology with the largest evaluated likelihood value is selected as the system topology, meaning that: 
\beq
\thet^*:=\thet_{i^*}, \text{ where }i^*=\argmax_{i=1,...,t+1} L_i\,.
\eeq
The proposed method's algorithm is described in full detail in Algorithm~\ref{alg:1}. The computation of the log-likelihood determines the complexity of the algorithm at each step of Bayesian optimization, which is of order $O(2^{2d}T)$.

\begin{figure}[ht!]
\centering
\includegraphics[width=87mm]{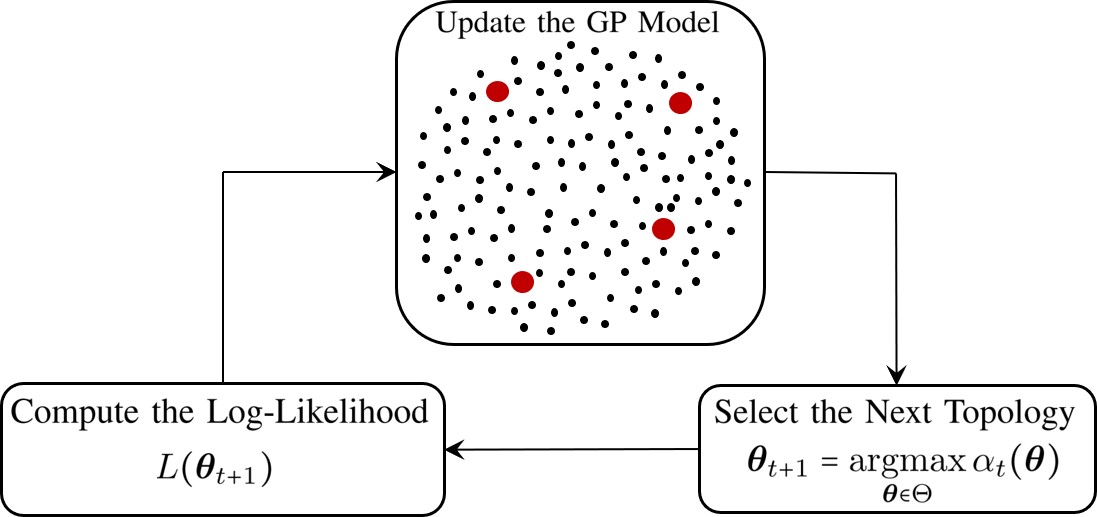}\vspace{0ex}
 \caption{Schematic diagram of the the proposed topology inference in GRNs.}
 \label{fig:schem}
\end{figure}

The inference process consists of three main components. Figure~\ref{fig:schem} represents the schematic diagram of the proposed method. The GP model predicts the log-likelihood values over the possible topology candidates, denoted by the black dots in Figure~\ref{fig:schem}. The red dots denote the evaluated log-likelihood values for the selected topologies up to the current iteration. Using the posterior distribution of the GP model, the next topology with the highest acquisition function is selected, followed by the log-likelihood evaluation for the selected topology. This sequential process continues until a stopping criterion is met.

\section{Numerical Experiments}
\begin{figure}[ht!]
\centerline{\includegraphics[width=85mm]{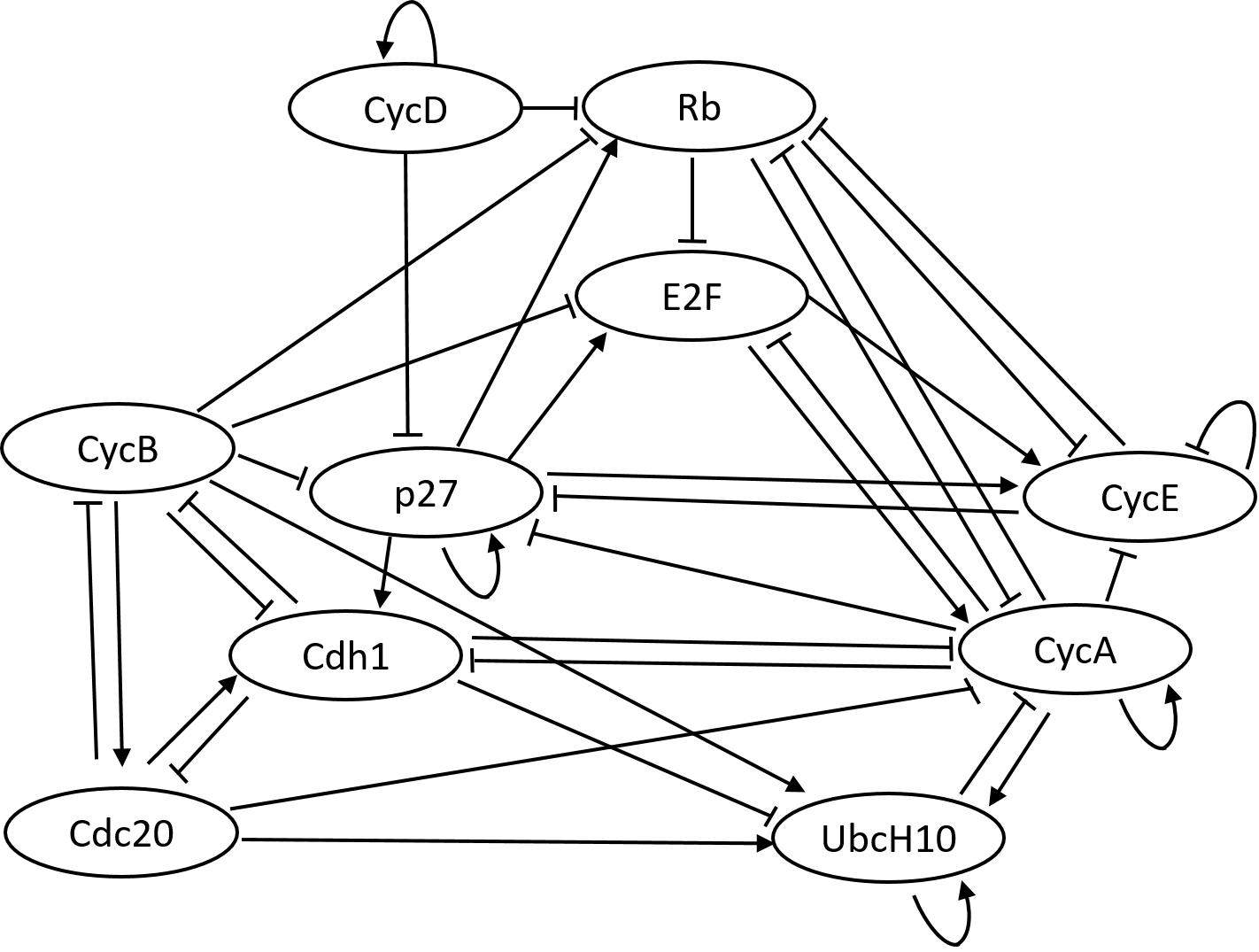}}\vspace{-.71ex}
\caption{Pathway diagram for the cell-cycle network.}
\label{fig:mammalian}
\end{figure}

The well-known mammalian cell-cycle network~\cite{faure2006dynamical} is used to evaluate the performance of our proposed method. Figure~\ref{fig:mammalian} presents the pathway diagram of this network. The state vector for this network is assumed as the following $\x = $(CycD, Rb, p27, E2F, CycE, CycA, Cdc20, Cdh1, UbcH10, CycB). The division of mammalian cells depends on the overall organism growth, controlled using signals that activate cyclin D (CycD) in the cell. 

\begin{figure*}[ht!]
\centerline{\includegraphics[width=175mm]{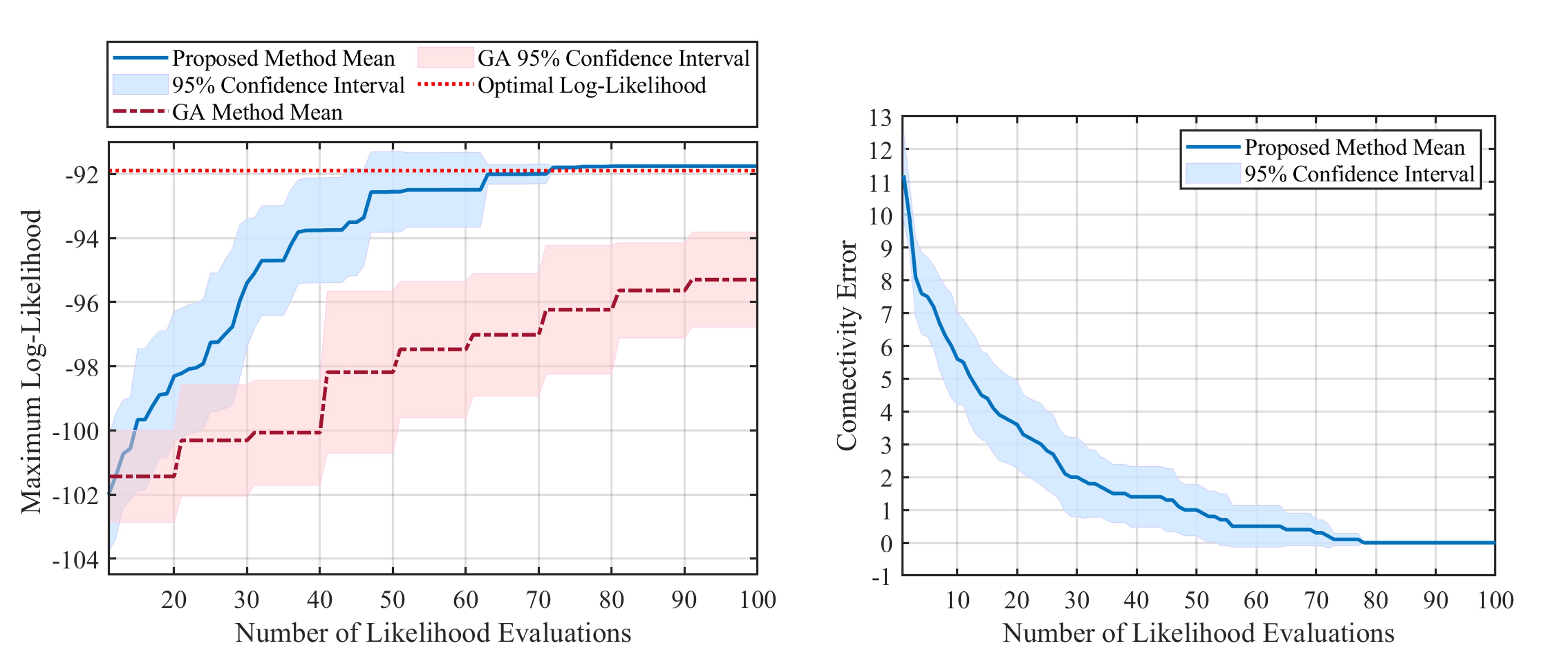}}\vspace{-.4ex}
\caption{Results of the mammalian cell-cycle network with $10$ unknown interactions.}
\label{fig:mammalian_1}
\end{figure*}
\vspace{0ex}

The connectivity matrix and bias vector in (\ref{eq:BN3}) for the mammalian cell-cycle network can be represented as:

{\small
\beq
\bal
{\bf C}&=\begin{bmatrix}
+1 & 0 & 0 & 0 & 0 & 0 & 0 & 0 & 0 & 0 \\
-1 & 0 & +1 & 0 & -1 & -1 & 0 & 0 & 0 & -1 \\
-1 & 0 & +1 & 0 & -1 & -1 & 0 & 0 & 0 & -1 \\
0 & -1 & +1 & 0 & 0 & -1 & 0 & 0 & 0 & -1 \\
0 & -1 & +1 & +1 & -1 & -1 & 0 & 0 & 0 & 0 \\
0 & -1 & 0 & +1 & 0 & +1 & -1 & -1 & -1 & 0 \\
0 & 0 & 0 & 0 & 0 & 0 & -1 & 0 & 0 & +1 \\
0 & 0 & +1 & 0 & 0 & -1 & +1 & 0 & 0 & -1 \\
0 & 0 & 0 & 0 & 0 & +1 & +1 & -1 & +1 & +1 \\
0 & 0 & 0 & 0 & 0 & 0 & -1 & -1 & 0 & 0 \\
\end{bmatrix},\\ 
{\bf b}&=
\begin{bmatrix} \frac{-1}{2} & \frac{-1}{2} & \frac{-1}{2} & \frac{-1}{2} & \frac{-1}{2} & \frac{-1}{2} & \frac{-1}{2} & \frac{-1}{2} & \frac{-1}{2} & \frac{-1}{2}\end{bmatrix}^T. 
\label{eq-newC}
\eal
\eeq}

In this section, we are assuming that the connectivity matrix is not fully known. This network has 10 genes, and there is a total of $2^{10}=1,024$ possible states for this network. Consequently, the transition matrix size is $2^{10}\times 2^{10}$, which causes the likelihood evaluation to be computationally expensive for any possible topology. Using our proposed method, we show that the optimal topology with the largest log-likelihood value can be inferred with few likelihood evaluations; hence, we offer an efficient search over all possible topologies.

In all of the experiments, 10 unknown interactions ($c_{ij}$) were considered. Each of the unknown interactions can take their values in the set $\{+1,0,-1\}$,  which leads to $3^{10}=59,049$ different possible system models, i.e., $\Theta=\{\thet^1,...,\thet^{3^{10}}\}$. The 10 randomly chosen unknown regulations, which are elements of the connectivity matrix in~(\ref{eq-newC}), are: 
\beq
\bal
& c_{2\,1} = -1, \quad  c_{3\,5} = -1, \quad c_{3\,10} = -1, \quad c_{4\,2}= -1, \quad  c_{5\,4}= +1 \\
& c_{6\,7} = -1, \quad  c_{6\,9} = -1, \quad c_{8\,3} = +1, \quad c_{9\,6}= +1, \quad c_{9\,8}= -1.
\eal
\label{eq-aij2}
\eeq
 The parameters used throughout the numerical experiments are expressed in Table~\ref{table:para2}. We also considered a uniform prior distribution for the initial states, i.e.,  $\PI^{\thet}_{0|0}(i)=\frac{1}{2^{10}}$ for all $\thet\in\Theta$ and $i=1,2,...,2^{10}$. Furthermore, all the experiments are repeated for 10 independent runs, and the average results along the confidence bounds are reported in all the figures. Note that the randomness of early results come from the process noise ($p$), and the way the sequential topology optimization is being performed in each run.


\begin{table}[ht!]
\centering
\caption{Parameter values of mammalian cell-cycle network experiments.}
  \small
\begin{tabular}{l*{1}{cc}r}
\midrule
\hspace{.5ex}\textbf{Parameter} & \hspace{-0ex}\textbf{Value}\\
\midrule
\hspace{.5ex}\text{Trajectory Length, $k$}  & \hspace{-0ex}  $100$ \\
\midrule
\hspace{.5ex}\text{Number of Likelihood Evaluations}  & \hspace{-0ex}  $100$ \\
\midrule
\hspace{.5ex}\text{Number of Genes, $d$} & \hspace{-0ex} $10$\\
\midrule
\hspace{.5ex}\text{Number of Unknown Regulations, $L$} & \hspace{-0ex} $10$\\
\midrule
\hspace{.5ex}\text{Process Noise, $p$} & \hspace{-0ex} $0.1$\\
\midrule
\hspace{.5ex}\text{Missing Data Percentage} & \hspace{-0ex} $50\%$\\
\midrule 
\end{tabular}
\label{table:para2}
\end{table}

For the first set of experiments, the performance of the proposed method is shown using two plots in Figure \ref{fig:mammalian_1}. The left plot represents the progress of the log-likelihood value of the inferred model with respect to the number of likelihood evaluations, meaning that it shows the maximum log-likelihood value obtained during the optimization process. Larger log-likelihood values mean that the chosen model can better represent the true model (i.e., the available data is more likely to come from models with larger likelihood values). As a comparison, we also repeated the same experiment using Genetic Algorithm (GA)~\cite{anderson1994genetic,whitley1994genetic}, which is a powerful and well-known solver for non-continuous problems. By looking at the left plot in Figure \ref{fig:mammalian_1}, we can see that the inference by the proposed method, indicated by the solid blue line, is better than the GA method (dashed red line). This superiority can be seen in terms of the mean and confidence intervals in Figure~\ref{fig:mammalian_1}. As we evaluate more likelihoods for different models, the likelihood of the proposed method's inferred model gets closer to the optimal log-likelihood value, indicated by the dotted red line. Hence, our proposed method is capable of reaching a better log-likelihood with less number of likelihood evaluations and has a more efficient way of searching over all the possible models. Furthermore, the 95\% confidence interval is illustrated in the same plot for both methods during this experiment. We can observe that the proposed method's confidence interval keeps getting smaller, and roughly after 70 evaluations, the confidence interval tends to go zero. This indicates the robustness of the proposed method, where after roughly 70 iterations, the log-likelihood gets to its optimal value at different independent runs. By contrast, the results from the GA still show a large confidence interval even after 100 evaluations, and its average is far less than the optimal log-likelihood value.  

The right plot of Figure \ref{fig:mammalian_1} shows the progress of the connectivity error during the optimization process (i.e., number of likelihood evaluations) obtained by the proposed method. Let $\C^{*}$ be the vectorized true connectivity matrix indicated in (\ref{eq-newC}), and $\C^{t}$ be the vectorized inferred connectivity matrix at $t$th likelihood evaluation. The connectivity error at iteration $t$ is defined as $||\C^{*}-\C^{t}||_{1}$. Evidently, we will have a better estimate of the true model as this error gets closer to zero. In the right plot, we can see that the connectivity error decreases as we do more evaluations, and after about 75 likelihood evaluations, the error gets to zero, meaning that we successfully inferred the true connectivity matrix. Also, as expected, we can see that the 95\% confidence interval gets smaller as we do more evaluations and eventually gets close to zero after about 75 evaluations. 

\begin{figure}[ht!]
\centerline{\includegraphics[width=85mm]{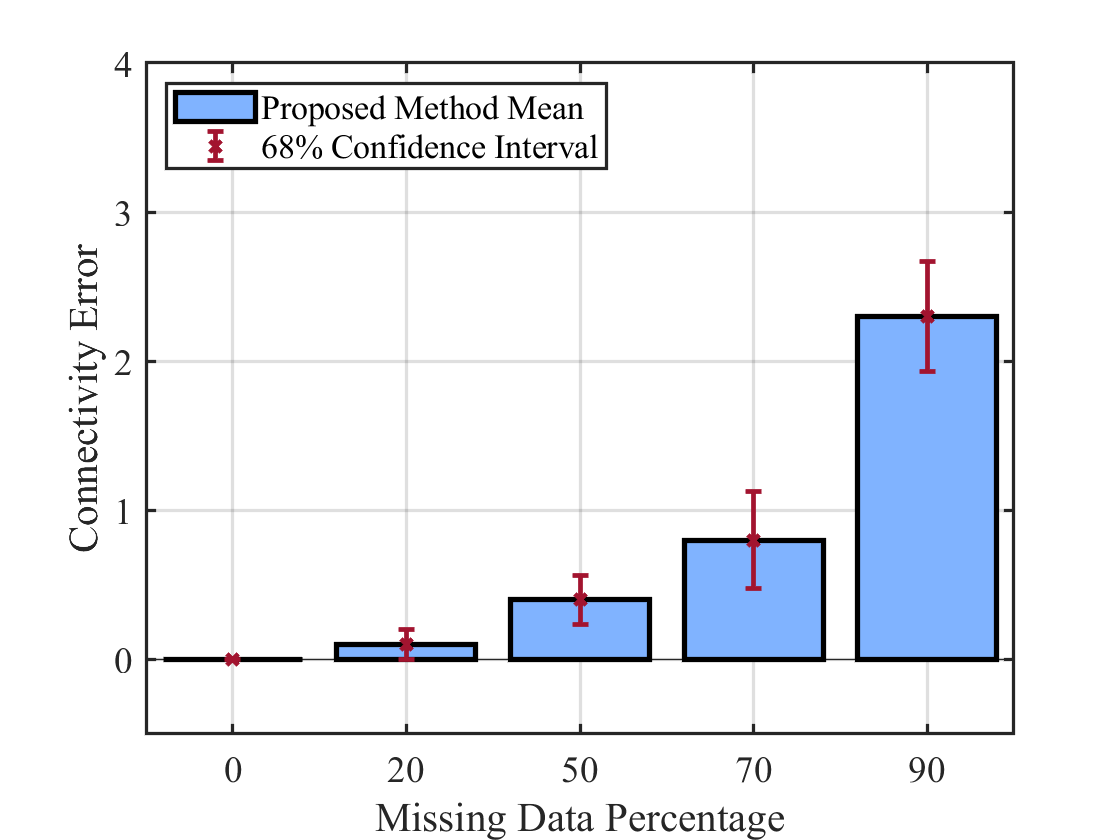}}\vspace{-.71ex}
\caption{Performance of the proposed method with respect to percentage of missing data.}
\label{fig:mammalian_2}
\end{figure}

In the second set of experiments, we aim to investigate the effect of missing data percentage on the performance of the proposed method. It is expected that with more missing data, it would be more difficult to infer the relationship between different components of the system; hence the connectivity error for the inferred model would be larger. For these experiments, we changed the missing data percentage from 0\% to 90\% and used Bernoulli noise value 0.2. Other parameters are fixed based on Table \ref{table:para2}. The mean of the inferred models' connectivity error obtained from these experiments, along with their 68\% confidence interval are presented as bar plots in Figure \ref{fig:mammalian_2}. As expected, these results demonstrate that the mean of connectivity error increases as the missing data percentage gets larger.  

\begin{figure}[ht!]
\centerline{\includegraphics[width=85mm]{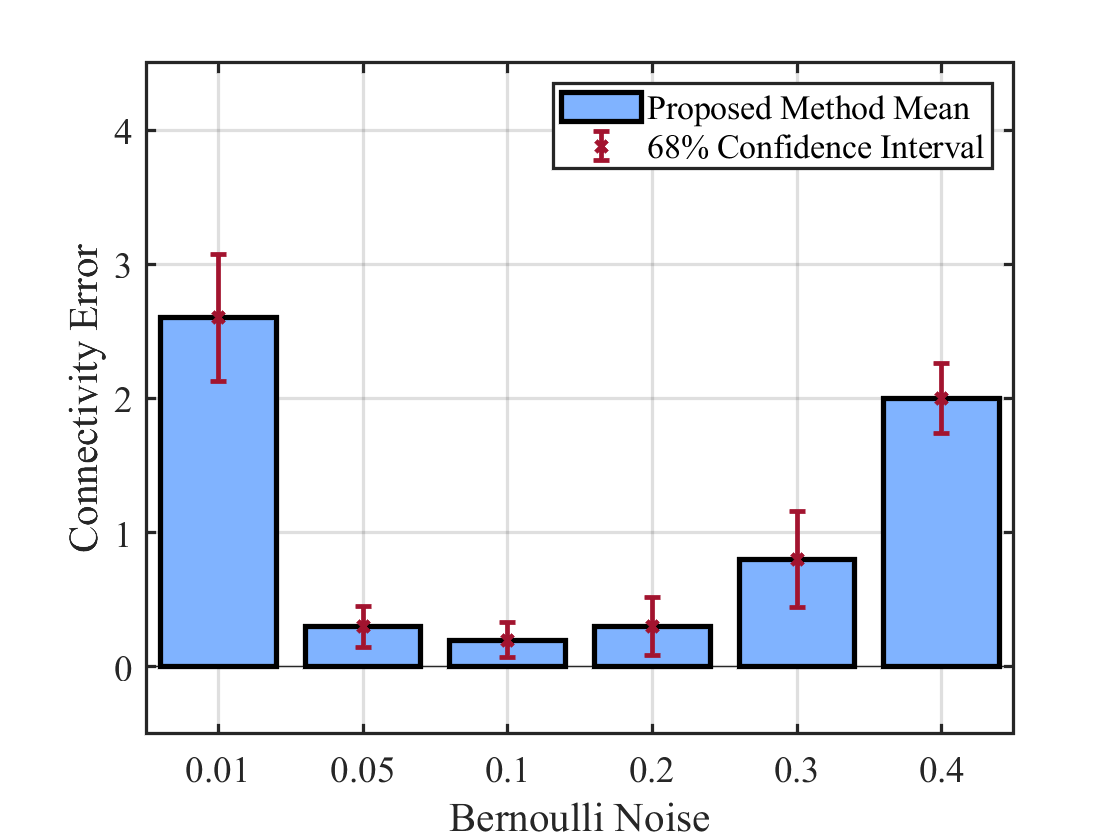}}\vspace{-.71ex}
\caption{Performance of the proposed method in presence of different Bernoulli noise.}
\label{fig:mammalian_3}
\end{figure}

The final set of experiments focuses on how the Bernoulli noise affects the performance of the proposed method. In all of these experiments, we consider 50\% missing data percentage, and we change the Bernoulli noise from 0.01 to 0.4. For performance comparison, the mean of the inferred models' connectivity error derived from these experiments is shown using bar plots in Figure \ref{fig:mammalian_3}. In this bar plot, we can observe that the connectivity error is large for the Bernoulli noise 0.01. As we increase the noise to 0.05 and 0.1, the connectivity error keeps decreasing. However, increasing the noise to 0.2, 0.3, and finally 0.4 results in a continuous increase in the connectivity error. This clearly shows the effect of stochasticity on inference. When the noise is very small (0.01), there are not many changes in the states of the system, hence, making it more difficult to infer different relations between components of the system. As we increase the stochasticity to some extent (0.1 in this experiment), there will be more transitions between the states, which makes the inference easier, and the connectivity error keeps getting smaller. However, if we keep increasing the noise (in this case from 0.2 to 0.4), the state transitions become more chaotic, which can make it more difficult to infer the true relations between the components and might even sometimes cause a wrong inference. Consequently, too much stochasticity can negatively affect the inference and might result in a large connectivity error mean. 

\section{Conclusion}
This paper presents a highly scalable topology inference method for gene regulatory networks (GRNs)  observed through temporally sparse data. The Boolean network model is used for capturing the dynamics of the GRNs. The inference process consists of inferring the interactions between genes or equivalently selecting a topology for the system among all the possible topologies that have the highest likelihood value. Evaluating the likelihood function for any given topology is expensive, preventing exhaustive search over the large possible topology space. The proposed method models the log-likelihood function by a Gaussian Process (GP) model with a structurally-inspired kernel function. This GP model captures the correlation between different possible topologies and provides the Bayesian representation of the log-likelihood function. Using the posterior distribution of the GP model, Bayesian optimization is used to efficiently search over the topology space. 

The high performance of our proposed method is shown using multiple experiments on the well-known mammalian cell-cycle network. We have also repeated all the experiments multiple times to obtain a confidence interval and further demonstrate the accuracy and robustness of the solutions obtained by our method. In the first experiment, we considered the topology inference of the mammalian cell-cycle network with 10 unknown interactions and 50\% missing data. From comparing the results of topology inference using our proposed method and genetic algorithm, we observed that our method is more efficient in searching over the topology space and reaches an optimal model with fewer likelihood evaluations. Meanwhile, the small confidence interval of our method justified the robustness of the solutions. The second experiment investigated the effect of missing data on the performance of the proposed inference method. From the results, we understand that as expected, with more missing data, the method's accuracy reduces, and the inference error becomes larger. Finally, in the third experiment, we studied the performance of our method in the presence of different Bernoulli noise (i.e., stochasticity in the state process). The results show that for small stochasticity, the accuracy of the inference is low, as the system spends most of its time in a few states (i.e., attractors) and the interactions between different components of the system are not distinguishable. As the stochasticity increases, the accuracy of the proposed method increases (as the error decreases) until a certain point, and after that again, the accuracy starts decreasing. This is because too much stochasticity turns the system into a more chaotic form, making the inference of the true model more challenging. 

\begin{acks}
This work has been supported by the National Institutes of Health award 1R21EB032480-01, National Science Foundation award IIS-2202395 and ARMY Research Office award W911NF2110299. 
\end{acks}

\bibliographystyle{ACM-Reference-Format}
\bibliography{ArxivSubmission.bib}


\begin{thebibliography}{40}


\ifx \showCODEN    \undefined \def \showCODEN     #1{\unskip}     \fi
\ifx \showDOI      \undefined \def \showDOI       #1{#1}\fi
\ifx \showISBNx    \undefined \def \showISBNx     #1{\unskip}     \fi
\ifx \showISBNxiii \undefined \def \showISBNxiii  #1{\unskip}     \fi
\ifx \showISSN     \undefined \def \showISSN      #1{\unskip}     \fi
\ifx \showLCCN     \undefined \def \showLCCN      #1{\unskip}     \fi
\ifx \shownote     \undefined \def \shownote      #1{#1}          \fi
\ifx \showarticletitle \undefined \def \showarticletitle #1{#1}   \fi
\ifx \showURL      \undefined \def \showURL       {\relax}        \fi
\providecommand\bibfield[2]{#2}
\providecommand\bibinfo[2]{#2}
\providecommand\natexlab[1]{#1}
\providecommand\showeprint[2][]{arXiv:#2}

\bibitem[Aalto et~al\mbox{.}(2020)]%
        {BMC3}
\bibfield{author}{\bibinfo{person}{Atte Aalto}, \bibinfo{person}{Lauri
  Viitasaari}, \bibinfo{person}{Pauliina Ilmonen}, \bibinfo{person}{Laurent
  Mombaerts}, {and} \bibinfo{person}{Jorge Goncalves}.}
  \bibinfo{year}{2020}\natexlab{}.
\newblock \showarticletitle{Gene regulatory network inference from sparsely
  sampled noisy data}.
\newblock \bibinfo{journal}{\emph{Nature Communications}}  \bibinfo{volume}{11}
  (\bibinfo{date}{07} \bibinfo{year}{2020}).
\newblock


\bibitem[Anderson and Ferris(1994)]%
        {anderson1994genetic}
\bibfield{author}{\bibinfo{person}{Edward~J Anderson} {and}
  \bibinfo{person}{Michael~C Ferris}.} \bibinfo{year}{1994}\natexlab{}.
\newblock \showarticletitle{Genetic algorithms for combinatorial optimization:
  the assemble line balancing problem}.
\newblock \bibinfo{journal}{\emph{ORSA Journal on Computing}}
  \bibinfo{volume}{6}, \bibinfo{number}{2} (\bibinfo{year}{1994}),
  \bibinfo{pages}{161--173}.
\newblock


\bibitem[Atabakhsh(1991)]%
        {atabakhsh1991survey}
\bibfield{author}{\bibinfo{person}{Homa Atabakhsh}.}
  \bibinfo{year}{1991}\natexlab{}.
\newblock \showarticletitle{A survey of constraint based scheduling systems
  using an artificial intelligence approach}.
\newblock \bibinfo{journal}{\emph{Artificial Intelligence in Engineering}}
  \bibinfo{volume}{6}, \bibinfo{number}{2} (\bibinfo{year}{1991}),
  \bibinfo{pages}{58--73}.
\newblock


\bibitem[Barman and Kwon(2020)]%
        {BMC2}
\bibfield{author}{\bibinfo{person}{Shohag Barman} {and}
  \bibinfo{person}{Yung-Keun Kwon}.} \bibinfo{year}{2020}\natexlab{}.
\newblock \showarticletitle{{A neuro-evolution approach to infer a Boolean
  network from time-series gene expressions}}.
\newblock \bibinfo{journal}{\emph{Bioinformatics}}  \bibinfo{volume}{36}
  (\bibinfo{date}{12} \bibinfo{year}{2020}), \bibinfo{pages}{i762--i769}.
\newblock
\showISSN{1367-4803}


\bibitem[Chan et~al\mbox{.}(2017)]%
        {chan2017gene}
\bibfield{author}{\bibinfo{person}{Thalia~E Chan}, \bibinfo{person}{Michael~PH
  Stumpf}, {and} \bibinfo{person}{Ann~C Babtie}.}
  \bibinfo{year}{2017}\natexlab{}.
\newblock \showarticletitle{Gene regulatory network inference from single-cell
  data using multivariate information measures}.
\newblock \bibinfo{journal}{\emph{Cell systems}} \bibinfo{volume}{5},
  \bibinfo{number}{3} (\bibinfo{year}{2017}), \bibinfo{pages}{251--267}.
\newblock


\bibitem[Chiquet et~al\mbox{.}(2019)]%
        {chiquet2019multiattribute}
\bibfield{author}{\bibinfo{person}{Julien Chiquet}, \bibinfo{person}{Guillem
  Rigaill}, {and} \bibinfo{person}{Martina Sundqvist}.}
  \bibinfo{year}{2019}\natexlab{}.
\newblock \bibinfo{booktitle}{\emph{A multiattribute Gaussian graphical model
  for inferring multiscale regulatory networks: an application in breast
  cancer}}.
\newblock \bibinfo{publisher}{Springer New York}, \bibinfo{address}{New York,
  NY}, \bibinfo{pages}{143--160}.
\newblock


\bibitem[de~Luis~Balaguer and Sozzani(2017)]%
        {de2017inferring}
\bibfield{author}{\bibinfo{person}{Maria~Angels de Luis~Balaguer} {and}
  \bibinfo{person}{Rosangela Sozzani}.} \bibinfo{year}{2017}\natexlab{}.
\newblock \bibinfo{booktitle}{\emph{Inferring gene regulatory networks in the
  arabidopsis root using a dynamic Bayesian network approach}}.
\newblock \bibinfo{publisher}{Springer New York}, \bibinfo{address}{New York,
  NY}, \bibinfo{pages}{331--348}.
\newblock


\bibitem[Dong et~al\mbox{.}(2013)]%
        {dong2013inference}
\bibfield{author}{\bibinfo{person}{Zijian Dong}, \bibinfo{person}{Tiecheng
  Song}, {and} \bibinfo{person}{Chuang Yuan}.} \bibinfo{year}{2013}\natexlab{}.
\newblock \showarticletitle{Inference of gene regulatory networks from genetic
  perturbations with linear regression model}.
\newblock \bibinfo{journal}{\emph{PloS one}} \bibinfo{volume}{8},
  \bibinfo{number}{12} (\bibinfo{year}{2013}), \bibinfo{pages}{e83263}.
\newblock


\bibitem[Faur{\'e} et~al\mbox{.}(2006)]%
        {faure2006dynamical}
\bibfield{author}{\bibinfo{person}{Adrien Faur{\'e}},
  \bibinfo{person}{Aur{\'e}lien Naldi}, \bibinfo{person}{Claudine Chaouiya},
  {and} \bibinfo{person}{Denis Thieffry}.} \bibinfo{year}{2006}\natexlab{}.
\newblock \showarticletitle{Dynamical analysis of a generic {B}oolean model for
  the control of the mammalian cell cycle}.
\newblock \bibinfo{journal}{\emph{Bioinformatics}} \bibinfo{volume}{22},
  \bibinfo{number}{14} (\bibinfo{year}{2006}), \bibinfo{pages}{e124--e131}.
\newblock


\bibitem[Gao et~al\mbox{.}(2018)]%
        {BMC1}
\bibfield{author}{\bibinfo{person}{Shuhua Gao}, \bibinfo{person}{Cheng Xiang},
  \bibinfo{person}{Changkai Sun}, \bibinfo{person}{Kairong Qin}, {and}
  \bibinfo{person}{Tong heng Lee}.} \bibinfo{year}{2018}\natexlab{}.
\newblock \showarticletitle{Efficient Boolean Modeling of Gene Regulatory
  Networks via random forest based feature selection and best-fit extension}.
\newblock \bibinfo{journal}{\emph{2018 IEEE 14th International Conference on
  Control and Automation (ICCA)}} (\bibinfo{year}{2018}),
  \bibinfo{pages}{1076--1081}.
\newblock


\bibitem[Godsill et~al\mbox{.}(2004)]%
        {Godsill2004}
\bibfield{author}{\bibinfo{person}{Simon~J Godsill}, \bibinfo{person}{Arnaud
  Doucet}, {and} \bibinfo{person}{Mike West}.} \bibinfo{year}{2004}\natexlab{}.
\newblock \showarticletitle{Monte {C}arlo smoothing for nonlinear time series}.
\newblock \bibinfo{journal}{\emph{Journal of the american statistical
  association}} \bibinfo{volume}{99}, \bibinfo{number}{465}
  (\bibinfo{year}{2004}), \bibinfo{pages}{156--168}.
\newblock


\bibitem[Hajiramezanali et~al\mbox{.}(2019)]%
        {EhsanBMCGenomics}
\bibfield{author}{\bibinfo{person}{Ehsan Hajiramezanali},
  \bibinfo{person}{Mahdi Imani}, \bibinfo{person}{Ulisses Braga-Neto},
  \bibinfo{person}{Xiaoning Qian}, {and} \bibinfo{person}{Edward~R Dougherty}.}
  \bibinfo{year}{2019}\natexlab{}.
\newblock \showarticletitle{Scalable optimal {B}ayesian classification of
  single-cell trajectories under regulatory model uncertainty}.
\newblock \bibinfo{journal}{\emph{BMC genomics}} \bibinfo{volume}{20},
  \bibinfo{number}{S6} (\bibinfo{year}{2019}).
\newblock


\bibitem[Henriques et~al\mbox{.}(2015)]%
        {henriques2015reverse}
\bibfield{author}{\bibinfo{person}{David Henriques}, \bibinfo{person}{Miguel
  Rocha}, \bibinfo{person}{Julio Saez-Rodriguez}, {and}
  \bibinfo{person}{Julio~R Banga}.} \bibinfo{year}{2015}\natexlab{}.
\newblock \showarticletitle{Reverse engineering of logic-based differential
  equation models using a mixed-integer dynamic optimization approach}.
\newblock \bibinfo{journal}{\emph{Bioinformatics}} \bibinfo{volume}{31},
  \bibinfo{number}{18} (\bibinfo{year}{2015}), \bibinfo{pages}{2999--3007}.
\newblock


\bibitem[H{\"u}rzeler and K{\"u}nsch(1998)]%
        {Hurzeler1998}
\bibfield{author}{\bibinfo{person}{Markus H{\"u}rzeler} {and}
  \bibinfo{person}{Hans~R K{\"u}nsch}.} \bibinfo{year}{1998}\natexlab{}.
\newblock \showarticletitle{Monte {C}arlo approximations for general
  state-space models}.
\newblock \bibinfo{journal}{\emph{Journal of Computational and Graphical
  Statistics}} \bibinfo{volume}{7}, \bibinfo{number}{2} (\bibinfo{year}{1998}),
  \bibinfo{pages}{175--193}.
\newblock


\bibitem[Imani and Braga-Neto(2017)]%
        {ImanBrag:TSPJ2}
\bibfield{author}{\bibinfo{person}{Mahdi Imani} {and}
  \bibinfo{person}{Ulisses~M Braga-Neto}.} \bibinfo{year}{2017}\natexlab{}.
\newblock \showarticletitle{Maximum-likelihood adaptive filter for partially
  observed {B}oolean dynamical systems}.
\newblock \bibinfo{journal}{\emph{IEEE Transactions on Signal Processing}}
  \bibinfo{volume}{65}, \bibinfo{number}{2} (\bibinfo{year}{2017}),
  \bibinfo{pages}{359--371}.
\newblock


\bibitem[Imani et~al\mbox{.}(2018)]%
        {Iman:CIJ5}
\bibfield{author}{\bibinfo{person}{Mahdi Imani}, \bibinfo{person}{Roozbeh
  Dehghannasiri}, \bibinfo{person}{Ulisses~M Braga-Neto}, {and}
  \bibinfo{person}{Edward~R Dougherty}.} \bibinfo{year}{2018}\natexlab{}.
\newblock \showarticletitle{Sequential experimental design for optimal
  structural intervention in gene regulatory networks based on the mean
  objective cost of uncertainty}.
\newblock \bibinfo{journal}{\emph{Cancer informatics}}  \bibinfo{volume}{17}
  (\bibinfo{year}{2018}), \bibinfo{pages}{1176935118790247}.
\newblock


\bibitem[Imani et~al\mbox{.}(2020)]%
        {ImanBrag:BBKFJ12}
\bibfield{author}{\bibinfo{person}{Mahdi Imani}, \bibinfo{person}{Edward
  Dougherty}, {and} \bibinfo{person}{Ulisses Braga-Neto}.}
  \bibinfo{year}{2020}\natexlab{}.
\newblock \showarticletitle{Boolean {K}alman Filter and Smoother Under Model
  Uncertainty}.
\newblock \bibinfo{journal}{\emph{Automatica}}  \bibinfo{volume}{111}
  (\bibinfo{year}{2020}), \bibinfo{pages}{108609}.
\newblock


\bibitem[Imani and Ghoreishi(2021)]%
        {ImanGh:TNNLSJ17}
\bibfield{author}{\bibinfo{person}{Mahdi Imani} {and}
  \bibinfo{person}{Seyede~Fatemeh Ghoreishi}.} \bibinfo{year}{2021}\natexlab{}.
\newblock \showarticletitle{Two-Stage {B}ayesian Optimization for Scalable
  Inference in State Space Models}.
\newblock \bibinfo{journal}{\emph{IEEE transactions on neural networks and
  learning systems}} (\bibinfo{year}{2021}).
\newblock


\bibitem[Imani et~al\mbox{.}(2019)]%
        {Iman:AAAIC15}
\bibfield{author}{\bibinfo{person}{Mahdi Imani},
  \bibinfo{person}{Seyede~Fatemeh Ghoreishi}, \bibinfo{person}{Douglas
  Allaire}, {and} \bibinfo{person}{Ulisses Braga-Neto}.}
  \bibinfo{year}{2019}\natexlab{}.
\newblock \showarticletitle{{MFBO-SSM}: {M}ulti-fidelity {B}ayesian
  optimization for fast inference in state-space models}.
\newblock \bibinfo{journal}{\emph{Proceedings of the AAAI Conference on
  Artificial Intelligence}} \bibinfo{volume}{33}, \bibinfo{number}{01}
  (\bibinfo{date}{Jul.} \bibinfo{year}{2019}), \bibinfo{pages}{7858--7865}.
\newblock


\bibitem[Johansen et~al\mbox{.}(2008)]%
        {johansen2008particle}
\bibfield{author}{\bibinfo{person}{Adam~M Johansen}, \bibinfo{person}{Arnaud
  Doucet}, {and} \bibinfo{person}{Manuel Davy}.}
  \bibinfo{year}{2008}\natexlab{}.
\newblock \showarticletitle{Particle methods for maximum likelihood estimation
  in latent variable models}.
\newblock \bibinfo{journal}{\emph{Statistics and Computing}}
  \bibinfo{volume}{18}, \bibinfo{number}{1} (\bibinfo{year}{2008}),
  \bibinfo{pages}{47--57}.
\newblock


\bibitem[Jones et~al\mbox{.}(1998)]%
        {jones1998efficient}
\bibfield{author}{\bibinfo{person}{Donald~R Jones}, \bibinfo{person}{Matthias
  Schonlau}, {and} \bibinfo{person}{William~J Welch}.}
  \bibinfo{year}{1998}\natexlab{}.
\newblock \showarticletitle{Efficient global optimization of expensive
  black-box functions}.
\newblock \bibinfo{journal}{\emph{Journal of Global optimization}}
  \bibinfo{volume}{13}, \bibinfo{number}{4} (\bibinfo{year}{1998}),
  \bibinfo{pages}{455--492}.
\newblock


\bibitem[Kantas et~al\mbox{.}(2015)]%
        {kantas2015particle}
\bibfield{author}{\bibinfo{person}{Nikolas Kantas}, \bibinfo{person}{Arnaud
  Doucet}, \bibinfo{person}{Sumeetpal~S Singh}, \bibinfo{person}{Jan
  Maciejowski}, {and} \bibinfo{person}{Nicolas Chopin}.}
  \bibinfo{year}{2015}\natexlab{}.
\newblock \showarticletitle{On particle methods for parameter estimation in
  state-space models}.
\newblock \bibinfo{journal}{\emph{Statistical science}} \bibinfo{volume}{30},
  \bibinfo{number}{3} (\bibinfo{year}{2015}), \bibinfo{pages}{328--351}.
\newblock


\bibitem[Kennedy and Eberhart(1995)]%
        {kennedy1995particle}
\bibfield{author}{\bibinfo{person}{James Kennedy} {and}
  \bibinfo{person}{Russell Eberhart}.} \bibinfo{year}{1995}\natexlab{}.
\newblock \showarticletitle{Particle swarm optimization}. In
  \bibinfo{booktitle}{\emph{Proceedings of ICNN'95-International Conference on
  Neural Networks}}, Vol.~\bibinfo{volume}{4}. IEEE,
  \bibinfo{pages}{1942--1948}.
\newblock


\bibitem[Kotiang and Eslami(2020)]%
        {kotiang2020probabilistic}
\bibfield{author}{\bibinfo{person}{Stephen Kotiang} {and} \bibinfo{person}{Ali
  Eslami}.} \bibinfo{year}{2020}\natexlab{}.
\newblock \showarticletitle{A probabilistic graphical model for system-wide
  analysis of gene regulatory networks}.
\newblock \bibinfo{journal}{\emph{Bioinformatics}} \bibinfo{volume}{36},
  \bibinfo{number}{10} (\bibinfo{year}{2020}), \bibinfo{pages}{3192--3199}.
\newblock


\bibitem[Langfelder and Horvath(2008)]%
        {langfelder2008wgcna}
\bibfield{author}{\bibinfo{person}{Peter Langfelder} {and}
  \bibinfo{person}{Steve Horvath}.} \bibinfo{year}{2008}\natexlab{}.
\newblock \showarticletitle{WGCNA: an R package for weighted correlation
  network analysis}.
\newblock \bibinfo{journal}{\emph{BMC bioinformatics}} \bibinfo{volume}{9},
  \bibinfo{number}{1} (\bibinfo{year}{2008}), \bibinfo{pages}{1--13}.
\newblock


\bibitem[Ma et~al\mbox{.}(2020)]%
        {ma2020inference}
\bibfield{author}{\bibinfo{person}{Baoshan Ma}, \bibinfo{person}{Mingkun Fang},
  {and} \bibinfo{person}{Xiangtian Jiao}.} \bibinfo{year}{2020}\natexlab{}.
\newblock \showarticletitle{Inference of gene regulatory networks based on
  nonlinear ordinary differential equations}.
\newblock \bibinfo{journal}{\emph{Bioinformatics}} \bibinfo{volume}{36},
  \bibinfo{number}{19} (\bibinfo{year}{2020}), \bibinfo{pages}{4885--4893}.
\newblock


\bibitem[Mandal et~al\mbox{.}(2016)]%
        {mandal2016reverse}
\bibfield{author}{\bibinfo{person}{Sudip Mandal}, \bibinfo{person}{Abhinandan
  Khan}, \bibinfo{person}{Goutam Saha}, {and} \bibinfo{person}{Rajat~Kumar
  Pal}.} \bibinfo{year}{2016}\natexlab{}.
\newblock \showarticletitle{Reverse engineering of gene regulatory networks
  based on S-systems and bat algorithm}.
\newblock \bibinfo{journal}{\emph{Journal of bioinformatics and computational
  biology}} \bibinfo{volume}{14}, \bibinfo{number}{03} (\bibinfo{year}{2016}),
  \bibinfo{pages}{1650010}.
\newblock


\bibitem[Mockus et~al\mbox{.}(1978)]%
        {mockus1978application}
\bibfield{author}{\bibinfo{person}{Jonas Mockus}, \bibinfo{person}{Vytautas
  Tiesis}, {and} \bibinfo{person}{Antanas Zilinskas}.}
  \bibinfo{year}{1978}\natexlab{}.
\newblock \showarticletitle{The application of {B}ayesian methods for seeking
  the extremum}.
\newblock \bibinfo{journal}{\emph{Towards global optimization}}
  \bibinfo{volume}{2}, \bibinfo{number}{117-129} (\bibinfo{year}{1978}),
  \bibinfo{pages}{2}.
\newblock


\bibitem[Ostrowski et~al\mbox{.}(2016)]%
        {ostrowski2016boolean}
\bibfield{author}{\bibinfo{person}{Max Ostrowski}, \bibinfo{person}{Lo{\"\i}c
  Paulev{\'e}}, \bibinfo{person}{Torsten Schaub}, \bibinfo{person}{Anne
  Siegel}, {and} \bibinfo{person}{Carito Guziolowski}.}
  \bibinfo{year}{2016}\natexlab{}.
\newblock \showarticletitle{Boolean network identification from perturbation
  time series data combining dynamics abstraction and logic programming}.
\newblock \bibinfo{journal}{\emph{Biosystems}}  \bibinfo{volume}{149}
  (\bibinfo{year}{2016}), \bibinfo{pages}{139--153}.
\newblock


\bibitem[Rasmussen and Williams(2006)]%
        {rasmussen2006gaussian}
\bibfield{author}{\bibinfo{person}{Carl~Edward Rasmussen} {and}
  \bibinfo{person}{Christopher Williams}.} \bibinfo{year}{2006}\natexlab{}.
\newblock \bibinfo{booktitle}{\emph{Gaussian processes for machine learning}}.
\newblock \bibinfo{publisher}{MIT Press}, \bibinfo{address}{Cambridge, United
  States}.
\newblock


\bibitem[Salleh et~al\mbox{.}(2017)]%
        {salleh2017multiple}
\bibfield{author}{\bibinfo{person}{Faridah Hani~Mohamed Salleh},
  \bibinfo{person}{Suhaila Zainudin}, {and} \bibinfo{person}{Shereena~M Arif}.}
  \bibinfo{year}{2017}\natexlab{}.
\newblock \showarticletitle{Multiple linear regression for reconstruction of
  gene regulatory networks in solving cascade error problems}.
\newblock \bibinfo{journal}{\emph{Advances in bioinformatics}}
  \bibinfo{volume}{2017} (\bibinfo{year}{2017}).
\newblock


\bibitem[Sch{\"o}n et~al\mbox{.}(2011)]%
        {Schon2011}
\bibfield{author}{\bibinfo{person}{Thomas~B Sch{\"o}n}, \bibinfo{person}{Adrian
  Wills}, {and} \bibinfo{person}{Brett Ninness}.}
  \bibinfo{year}{2011}\natexlab{}.
\newblock \showarticletitle{System identification of nonlinear state-space
  models}.
\newblock \bibinfo{journal}{\emph{Automatica}} \bibinfo{volume}{47},
  \bibinfo{number}{1} (\bibinfo{year}{2011}), \bibinfo{pages}{39--49}.
\newblock


\bibitem[Shmulevich and Dougherty(2010)]%
        {shmulevich2010probabilistic}
\bibfield{author}{\bibinfo{person}{Ilya Shmulevich} {and}
  \bibinfo{person}{Edward~R Dougherty}.} \bibinfo{year}{2010}\natexlab{}.
\newblock \bibinfo{booktitle}{\emph{Probabilistic {B}oolean networks: the
  modeling and control of gene regulatory networks}}.
  Vol.~\bibinfo{volume}{118}.
\newblock \bibinfo{publisher}{siam}, \bibinfo{address}{Philadelphia, United
  States}.
\newblock


\bibitem[Shmulevich et~al\mbox{.}(2002b)]%
        {shmulevich2002probabilistic}
\bibfield{author}{\bibinfo{person}{Ilya Shmulevich}, \bibinfo{person}{Edward~R
  Dougherty}, \bibinfo{person}{Seungchan Kim}, {and} \bibinfo{person}{Wei
  Zhang}.} \bibinfo{year}{2002}\natexlab{b}.
\newblock \showarticletitle{Probabilistic {B}oolean networks: a rule-based
  uncertainty model for gene regulatory networks}.
\newblock \bibinfo{journal}{\emph{Bioinformatics}} \bibinfo{volume}{18},
  \bibinfo{number}{2} (\bibinfo{year}{2002}), \bibinfo{pages}{261--274}.
\newblock


\bibitem[Shmulevich et~al\mbox{.}(2002a)]%
        {shmulevich2002boolean}
\bibfield{author}{\bibinfo{person}{Ilya Shmulevich}, \bibinfo{person}{Edward~R
  Dougherty}, {and} \bibinfo{person}{Wei Zhang}.}
  \bibinfo{year}{2002}\natexlab{a}.
\newblock \showarticletitle{From {B}oolean to probabilistic {B}oolean networks
  as models of genetic regulatory networks}.
\newblock \bibinfo{journal}{\emph{Proc. IEEE}} \bibinfo{volume}{90},
  \bibinfo{number}{11} (\bibinfo{year}{2002}), \bibinfo{pages}{1778--1792}.
\newblock


\bibitem[Villaverde et~al\mbox{.}(2013)]%
        {villaverde2013reverse}
\bibfield{author}{\bibinfo{person}{Alejandro~F Villaverde},
  \bibinfo{person}{John Ross}, {and} \bibinfo{person}{Julio~R Banga}.}
  \bibinfo{year}{2013}\natexlab{}.
\newblock \showarticletitle{Reverse engineering cellular networks with
  information theoretic methods}.
\newblock \bibinfo{journal}{\emph{Cells}} \bibinfo{volume}{2},
  \bibinfo{number}{2} (\bibinfo{year}{2013}), \bibinfo{pages}{306--329}.
\newblock


\bibitem[Vinciotti et~al\mbox{.}(2016)]%
        {vinciotti2016model}
\bibfield{author}{\bibinfo{person}{Veronica Vinciotti}, \bibinfo{person}{Luigi
  Augugliaro}, \bibinfo{person}{Antonino Abbruzzo}, {and}
  \bibinfo{person}{Ernst~C Wit}.} \bibinfo{year}{2016}\natexlab{}.
\newblock \showarticletitle{Model selection for factorial {G}aussian graphical
  models with an application to dynamic regulatory networks}.
\newblock \bibinfo{journal}{\emph{Statistical applications in genetics and
  molecular biology}} \bibinfo{volume}{15}, \bibinfo{number}{3}
  (\bibinfo{year}{2016}), \bibinfo{pages}{193--212}.
\newblock


\bibitem[Whitley(1994)]%
        {whitley1994genetic}
\bibfield{author}{\bibinfo{person}{Darrell Whitley}.}
  \bibinfo{year}{1994}\natexlab{}.
\newblock \showarticletitle{A genetic algorithm tutorial}.
\newblock \bibinfo{journal}{\emph{Statistics and computing}}
  \bibinfo{volume}{4}, \bibinfo{number}{2} (\bibinfo{year}{1994}),
  \bibinfo{pages}{65--85}.
\newblock


\bibitem[Wills et~al\mbox{.}(2013)]%
        {will2013}
\bibfield{author}{\bibinfo{person}{Adrian Wills}, \bibinfo{person}{Thomas~B
  Sch{\"o}n}, \bibinfo{person}{Lennart Ljung}, {and} \bibinfo{person}{Brett
  Ninness}.} \bibinfo{year}{2013}\natexlab{}.
\newblock \showarticletitle{Identification of hammerstein--wiener models}.
\newblock \bibinfo{journal}{\emph{Automatica}} \bibinfo{volume}{49},
  \bibinfo{number}{1} (\bibinfo{year}{2013}), \bibinfo{pages}{70--81}.
\newblock


\bibitem[Yang and Chen(2020)]%
        {yang2020overview}
\bibfield{author}{\bibinfo{person}{Bin Yang} {and} \bibinfo{person}{Yuehui
  Chen}.} \bibinfo{year}{2020}\natexlab{}.
\newblock \showarticletitle{Overview of gene regulatory network inference based
  on differential equation models}.
\newblock \bibinfo{journal}{\emph{Current Protein and Peptide Science}}
  \bibinfo{volume}{21}, \bibinfo{number}{11} (\bibinfo{year}{2020}),
  \bibinfo{pages}{1054--1059}.
\newblock


\end{thebibliography}










\end{document}